\documentclass[12pt,a4paper]{article}
\usepackage[top=2.2cm,bottom=2.5cm,left=2.2cm,right=2.2cm]{geometry}
\usepackage{latexsym,amsmath,amsfonts,amssymb,amsthm,amscd,mathrsfs}

\newcommand{\be}{\begin{equation}}
\newcommand{\ee}{\end{equation}}
\newcommand{\bea}{\begin{eqnarray}}
\newcommand{\eea}{\end{eqnarray}}

\newcommand{\ba}{\begin{aligned}}
\newcommand{\ea}{\end{aligned}}
\numberwithin{equation}{section}

\numberwithin{thmcounter}{section}

\theoremstyle{definition}
\newtheorem*{acknowledgements}{Acknowledgements}

\theoremstyle{plain}

\def\1{{\boldsymbol 1}}                     %
\def\cD{{\mathcal D}}                       %
\def\cG{{\mathcal G}}                       %
\def\cH{{\mathcal H}}                       %
\def\cP{{\mathcal P}}                       %
\def\tr{\mathrm{tr}}                        %
\def\diag{\mathrm{diag}}                    %
\def\ri{{\rm i}}                            %
\def\C{\mathbb{C}}                          %
\def\N{\mathbb{N}}                          %
\def\R{\mathbb{R}}                          %
\def\T{\mathbb{T}}                          %
\def\UN{{\rm U}}                            %
\def\GL{{\rm GL}}                           %
\def\fH{\mathfrak{H}}                       %
\def\cF{{\mathcal F}}                       %
\def\cM{{\mathcal M}}                       %
\def\cO{{\mathcal O}}                       %
\def\cT{{\mathcal T}}                       %
\def\reg{\mathrm{reg}}                      %
\def\red{\mathrm{red}}                      %
\def\ad{\mathrm{ad}}                        %
\def\cR{{\mathcal R}}                       %
\def\cA{{\mathcal A}}                       %
\def\dt {\left.\frac{d}{dt}\right|_{t=0}}   %
\def\gl{\mathrm{gl}}                        %
\def\u{\mathrm{u}}                          %
\def\ext{\mathrm{ext}}                      %
\def\herm{\mathrm{herm}}                    %
\def\V{\mathbb{V}}                          %
\def\pa{\partial}                           %
\def\fa{\mathfrak{A}}                       %
\def\Li{\mathrm{Li}}                        %
\def\cW{\mathcal{W}}                        %
\def\Der{\operatorname{\mathcal{D}}}        %
\def\cL{{L}}                                %
\begin{document}

\begin{center}
{\Large\bf
Bi-Hamiltonian structure of a dynamical system introduced by Braden and Hone}
\end{center}

\medskip
\begin{center}
L.~Feh\'er${}^{a,b}$
\\

\bigskip
${}^a$Department of Theoretical Physics, University of Szeged\\
Tisza Lajos krt 84-86, H-6720 Szeged, Hungary\\
e-mail: lfeher@physx.u-szeged.hu

\medskip
${}^b$Department of Theoretical Physics, WIGNER RCP, RMKI\\
H-1525 Budapest, P.O.B.~49, Hungary\\
\end{center}

\begin{abstract}
We investigate the finite dimensional dynamical system derived by Braden and Hone in 1996
from the solitons
of $A_{n-1}$ affine Toda field theory. This system of evolution equations for an $n\times n$ Hermitian matrix
$L$ and a real diagonal matrix $q$ with distinct eigenvalues was interpreted as a special
case of the spin Ruijsenaars--Schneider models due to Krichever and Zabrodin.
A decade later, L.-C. Li  re-derived the model from a general framework built on coboundary dynamical Poisson groupoids.
This led to a Hamiltonian description of the gauge invariant content of the model,
where the gauge transformations act as conjugations of $L$ by diagonal unitary matrices.
Here, we point out that the same dynamics can be interpreted also as a special case
of the spin Sutherland systems obtained by reducing the free geodesic motion on symmetric spaces,
studied by Pusztai and the author
in 2006;  the relevant symmetric space being $\mathrm{GL}(n,\mathbb{C})/ \mathrm{U}(n)$.
This construction provides an alternative Hamiltonian interpretation of the Braden--Hone dynamics.
We prove that the two Poisson brackets are compatible and yield
a bi-Hamiltonian description of the standard commuting flows of the model.
 \end{abstract}

{\linespread{0.8}\tableofcontents}

\newpage

\section{Introduction}

We are witnesses to intense recent interest in spin extensions \cite{GH,KZ,LX,Li1,Li2,FP1,FP2}
of the standard Calogero--Moser--Sutherland
and Ruijsenaars--Schneider type many-body models.
 Current studies \cite{Res, SS, KLOZ,ILLS, F, Fa, CF2, GG} are devoted to the mathematical
structure  and to interesting physical applications   of systems of this type.
In this paper we take a fresh look at a so far largely neglected, not yet well-understood, aspect
of such systems. Namely, we shall uncover a bi-Hamiltonian structure for a remarkable family of examples.

Let $L$ be an  $n\times n$ Hermitian matrix and $q=\diag(q_1,q_2,\dots, q_n)$ a real diagonal matrix
with distinct eigenvalues.
Braden and Hone \cite{BH} derived the following evolution equations from the affine Toda solitons:
\bea
&&\dot{q}_j = L_{jj}, \qquad
 \dot{L}_{jj} =2 \sum_{\ell \neq j} \vert L_{j\ell} \vert^2 \coth(q_j - q_\ell),
 \label{I1}\\
&&\dot{L}_{jk} =
 \sum_{\ell \neq j} L_{j\ell}  L_{\ell k} \coth(q_j - q_\ell) -
 \sum_{\ell \neq k}L_{j\ell}  L_{\ell k} \coth(q_\ell - q_k),\quad 1\leq j\neq k\leq n.
\nonumber \eea
In their context $L$ has a special form,
but the equations make sense for arbitrary $L$, and here we shall study
the system (\ref{I1}) in its general form.
It will be assumed that $q$ varies in the domain
\be
\cA^o:= \{ q \in \R^n\mid q_1> q_2 > \dots > q_n\}.
\label{I3}\ee
Note that $q\in \R^n$ and the corresponding diagonal matrix are denoted by the same letter.
The Braden--Hone equations represent the first member of a hierarchy \cite{Li2}, which
is conveniently described utilizing a dynamical $r$-matrix.

Let us consider the Lie algebra $\cG:= \u(n)$ and introduce $\cG^\C_\R:=
\gl(n,\C)$.  The notation emphasizes that we regard $\gl(n,\C)$
as a real Lie algebra. We equip
it with the invariant bilinear form
\be
\langle X, Y\rangle_\R:= \Re \tr(XY),\quad \forall X,Y\in \cG^\C_\R.
\label{I4}\ee
This induces the orthogonal vector space decomposition
\be
\cG^\C_\R = \cG + \ri \cG,
\label{I5}\ee
which can be further refined as
\be
\cG = \cT + \cT^\perp, \quad \ri \cG = \cA + \cA^\perp,
\label{I6}\ee
where $\cT$ (resp. $\cA$) consists of anti-Hermitian (resp. Hermitian) diagonal matrices,
and $\cT^\perp$ (resp. $\cA^\perp$) contains the corresponding off-diagonal matrices.
Taking any $w\in \cH^o\subset \cH$, where
\be
\cH:= \cA + \cT,\quad \cH^o := \cA^o + \cT,
\label{cH}\ee
we set
\be
\cR(w) X: = 0\,\,
\hbox{for}\,\, X\in (\cT + \cA), \,\,\hbox{and}\,\,
\cR(w)X:= (\coth \ad_w)(X) \,\,\hbox{for}\,\, X\in (\cT^\perp + \cA^\perp).
\label{R}\ee
This gives a well-defined linear operator on $\cG^\C_\R$ that
represents a solution
of the modified classical dynamical Yang--Baxter equation \cite{EV}.
By using this dynamical $r$-matrix, for any $m\in \N$, one can define the following system
of evolution equations:
\be
\dot{q_j} = (L^m)_{jj}, \quad \dot{L} = [\cR(q)L^m, L]
\quad \hbox{for}\quad (q,L)\in \cA^o\times \ri\cG.
\label{I8}\ee
For $m=1$, this is the Braden--Hone system (\ref{I1}).

An important feature of this system is that the evolutional derivations associated with different values of $m\in \N$
\emph{commute after restriction to gauge invariant functions}. By definition,
 a gauge invariant function $F$
of $q$ and $L$ satisfies
\be
 F(q, L)= F(q, \eta L \eta^{-1}) \qquad \forall \eta\in \T^n,
\ee
where $\T^n$ is the group of diagonal unitary matrices.
We introduce  the term `Braden--Hone hierarchy' to refer to the restrictions of the
derivations (\ref{I8}) to the gauge invariant functions.
The commutativity actually does not hold if we do not restrict to gauge invariant
functions. (See  Appendix A.)
This state of affairs hints that the Braden--Hone hierarchy should result from some suitable
Hamiltonian reduction,  for which the action of $\T^n$ should represent the gauge transformations on a moment map
`constraint surface'. It turns out that this expectation holds, and can be realized by (at least) two different
reduction procedures.

A reduction procedure leading to the Braden--Hone hierarchy
was found by L.-C. Li in \cite{Li2}, and another
one can be extracted with a little effort from the joint paper \cite{FP2}  by Pusztai
and the present author. The purpose of the current work is to show that the
Poisson brackets resulting from these two reduction procedures \emph{are compatible,
and equip the Braden--Hone hierarchy with a bi-Hamiltonian structure}.

Now we describe the two Poisson brackets and our main result.
For any real function $F\in C^\infty(\cA^o\times \ri \cG)$ we define its derivatives
$\nabla_1 F\in C^\infty(\cA^o \times \ri \cG,\cA)$ and $\nabla_2 F\in C^\infty(\cA^o \times \ri \cG,\ri \cG)$
 by requiring that at the point $(q,L)$ we have
\be
\langle \delta q, \nabla_1 F\rangle_\R = \dt F(q+ t \delta q, L)
\quad\hbox{and}\quad
\langle \delta L, \nabla_2 F\rangle_\R = \dt F(q, L + t \delta L)
\label{I10}\ee
for all $\delta q\in \cA$ and $\delta L \in \ri\cG$.
(Here, $F(q+ t \delta q, L)$ is well-defined for small $t$.)
For any $m\in \N$, let us define $H_m \in C^\infty(\cA^o \times \ri \cG)$ by
\be
H_m(q,L) := \frac{1}{m} \tr(L^m).
\label{I11}\ee
Let $V_m$ be the derivation\footnote{The action of $V_m$ on a  function $F$ is denoted $V_m[F]$;
the components of $q$ and $L$ are evaluation functions.}    of $C^\infty(\cA^o \times \ri \cG)$
generated by equation (\ref{I8}), i.e., by
the definition
\be
V_m[q_j]:= (L^m)_{jj}, \quad V_m[L]:= [\cR(q)L^m, L].
\label{Vm}\ee
By expanding $L$ in a basis $\{ Z^a\}$ of $\ri \cG$,  we have
$V_m[L_a Z^a] = V_m[L_a] Z^a$.
Note that $V_m$ maps  $C^\infty(\cA^o \times \ri \cG)^{\T^n}$,
the ring of gauge invariant functions,   to itself.

\medskip \noindent
{\bf Theorem 1.}  \emph{The following formulae define two Poisson brackets on $C^\infty(\cA^o\times \ri \cG)^{\T^n}$:
\be
\{ F,H\}_2(q,L) = \langle  \nabla_1 F, L \nabla_2 H\rangle_\R
- \langle  \nabla_1 H, L \nabla_2 F\rangle_\R
- 2\langle \cR(q)( L \nabla_2 F),  L \nabla_2 H \rangle_\R
\label{I13}\ee
and
\be
\{ F,H\}_1(q,L) = \langle  \nabla_1 F,  \nabla_2 H \rangle_\R
- \langle  \nabla_1 H,  \nabla_2 F\rangle_\R
+\langle L, [ \nabla_2 F,  \nabla_2 H]_{\cR(q)}   \rangle_\R,
\label{I14}\ee
where $[ X,Y]_{\cR(q)}:= [ \cR(q)X, Y] + [X, \cR(q)Y]$, and the $\nabla_i$ are taken at $(q,L)$.
The derivative of $F\in C^\infty(\cA^o\times \ri\cG)^{\T^n}$ along the vector field $V_m$  (\ref{Vm})
can be written in Hamiltonian form:
\be
V_m[F] = \{ F, H_m\}_2 = \{ F, H_{m+1}\}_1.
\label{I15}\ee
Moreover, we have $\{ H_\ell, H_m\}_2 = \{ H_\ell, H_m\}_1 =0$ for all $\ell, m\in \N$.}

\medskip
We shall explain that Theorem 1 follows by elaborating  earlier results found in \cite{Li2,FP2}.

Let $\Der$ be the derivation of the ring $C^\infty(\cA^o \times \ri\cG)$ defined by
\be
\Der[q_i]:=0, \quad \Der[L_{jk}] := \delta_{jk},
\label{Der}\ee
which preserves the gauge invariant functions. Our main result is then

\medskip\noindent
{\bf Theorem 2.}
\emph{The two Poisson brackets of Theorem 1 satisfy the relations
\be
\{ F, H\}_1 = \Der[\{F,H\}_2] - \{ \Der[F], H\}_2 - \{ F, \Der[H]\}_2,
\label{claim1}\ee
\be
\Der[\{ F,H\}_1] - \{ \Der[F], H\}_1 - \{ F, \Der[H]\}_1 =0.
\label{claim2}\ee
Consequently, they are compatible and define an exact bi-Hamiltonian structure. }

\medskip
The compatibility of the two Poisson brackets is a consequence of the relation (\ref{claim1}).
For readability, we quote the relevant well-known result together with an indication of its proof.

\medskip \noindent
{\bf Lemma 3.} \emph{Let $(\fa, \{\ ,\ \})$ be a Poisson algebra and $\Der$ a derivation
of the underlying commutative algebra $\fa$. Suppose that the  bracket
$\{ f, h\}^{\Der} := \Der[\{f,h\}] - \{\Der[f], h\} - \{ f, \Der[h]\}$
satisfies the Jacobi identity. Then the formula
\be
\{ f,h\}_{x,y} = x \{ f,h\} + y \{f,h\}^{\Der}
\label{xy}\ee
defines a Poisson bracket, for any constant parameters $x$ and $y$.}

\medskip\noindent
\begin{proof} For any derivation $\Der$, the bracket $\{\ ,\ \}_{x,y}$ is automatically
anti-symmetric and verifies the Leibniz property.
It is a simple exercise to verify the Jacobi identity
by direct calculation.
\end{proof}
The bi-Hamiltonian structures of the form (\ref{xy}) are called `exact'
when the application of $\Der$ to $\{\ ,\ \}^{\Der}$ gives zero, like in (\ref{claim2}).
Equation (\ref{I15}) and the compatibility of the two Poisson brackets together
show the bi-Hamiltonian character of the Braden--Hone hierarchy.
Throughout the paper,  we use the standard terminology of bi-Hamiltonian systems, see e.g.~\cite{FMP, Se, Sm},
although we are mostly dealing with Poisson algebras of gauge invariant functions, and not directly with
Poisson manifolds. We proceed in this manner in order to circumvent the complication
that the quotient space $(\cA^o \times \ri\cG)/\T^n$ is not a smooth manifold.
This should not lead to any confusion.

Now we sketch the organization of the text. Section 2 is devoted to a short summary
of the construction of the Poisson structure $\{\ ,\ \}_2$ due to L.-C. Li \cite{Li2}.
Our presentation contains some novel elements:
the precise connection to the original notations used in \cite{Li2} is described
in Appendix B. In Section 3 we expound the derivation of the Braden--Hone
hierarchy with its Poisson structure $\{\ ,\ \}_1$, building on the
paper \cite{FP2}. All results in Section 3 will be obtained relying on
analogous results of this reference, despite the fact that there reductions of geodesic motion on
\emph{simple} non-compact Lie groups were considered, which formally excludes our present case.
Besides explaining  Theorem 1, the goal of
Section  2 and Section 3 is to prepare
the ground for the proof of Theorem 2, which occupies Section 4.
Finally, we conclude in Section 5 by pointing out a few open problems for future work.

\section{The Poisson structure obtained by L.-C. Li}

We tersely summarize those points of the construction of \cite{Li2}, which are directly
relevant for us. Some details are relegated to Appendix B.

Let $G_\R^\C$ denote $\GL(n,\C)$ regarded as a real Lie group, and denote by
$\fH$ its closed submanifold consisting of the invertible Hermitian matrices of size $n$.
By applying a certain discrete reduction to a dynamical Poisson groupoid structure
on $\cH^o \times G^\C_\R\times \cH^o$, an interesting Poisson structure on
the manifold $\cH^o \times \fH$ was obtained.
This Poisson structure extends smoothly
from the dense open submanifold $\cH^o \times \fH \subset \cH^o \times \ri \cG$ to the
full of $\cH^o \times \ri\cG$.

Denote the elements of $\cH^o \times \ri \cG$ by pairs $(w,L)$. For any smooth
real function $F=F(w,L)$,
introduce the derivatives
$\nabla_1 F\in C^\infty(\cH^o \times \ri\cG, \cH)$
and $\nabla_2 F\in C^\infty(\cH\times \ri\cG, \ri\cG)$
 in complete analogy with the definition (\ref{I10}),
using the bilinear form (\ref{I4}).
Then, as is detailed in Appendix B, the Poisson structure given by equation (5.8) in \cite{Li2}
can be re-written in the following form:
\be
\{ F, H\}_\Li(w,L) =  \langle \nabla_1 F, L\nabla_2 H\rangle_\R - \langle \nabla_1 H, L \nabla_2 F\rangle_\R
-2 \langle \cR(w) (L \nabla_2F), L\nabla_2 H \rangle_\R .
\label{LiPB}\ee
Here, $F$ and $H$ are arbitrary elements of $C^\infty(\cH^o \times \ri\cG)$, the
derivatives are evaluated at $(w,L)$, and we use $\cR$ (\ref{R}).

For any $X\in \cT$, define  $w^X\in C^\infty(\cH^o \times \ri\cG)$ by $w^X(w,L):=\langle w, X\rangle_\R$.
The associated Hamiltonian vector field can be symbolically written as
\be
\{ w, w^X\}_\Li = 0,
\qquad
\{ L, w^X\}_\Li = -\frac{1}{2} [X,L].
\label{22}\ee
This encodes the Poisson brackets between $w^X$ and the matrix elements of $w$ and $L$,
regarded as functions on $\cH^o\times \ri \cG$.
The formula (\ref{22}) has the following important consequence:
\begin{itemize}
 \item{Take $\cT$ as the model of its own dual space by means of the trace pairing.
 Then the map $\phi: (w,L) \mapsto -2\Im(w)$ can be identified as the moment map
 for the Hamiltonian action of $\T^n$ whereby $\eta\in \T^n$ sends $(w,L)$ to $(w, \eta L \eta^{-1})$.}
\end{itemize}
Notice that the Hamiltonian $H_m\in C^\infty(\cH^o \times \ri \cG)$,
\be
H_m(w,L):= \frac{1}{m} \tr(L^m),
\label{Hmm}\ee is invariant under the above $\T^n$-action.
It follows that the Hamiltonian vector field
 generated by $H_m$ is tangent to the level surfaces
of the moment map $\phi$. The level surface $\phi=0$ is the submanifold
\be
\cA^o \times \ri\cG \subset \cH^o\times \ri\cG,
\ee
whose elements are denoted by pairs $(q,L)$, like in the Introduction.
An easy calculation gives that \emph{the restriction of the Hamiltonian vector field of $H_m$ (\ref{Hmm})  to this
level surface is precisely the vector field $V_m$ (\ref{Vm})}.
(It should not lead to any confusion that in equations (\ref{I11}) and (\ref{I15})
the corresponding restriction of $H_m$ (\ref{Hmm}) is denoted by the same letter.)

One knows from the general reduction theory (the theory of reduction
by first class constraints \`a la Dirac
is all what is needed here\footnote{We do not consider
`Dirac brackets', since our reductions do not admit globally valid gauge fixings.})
that $C^\infty(\cA^o \times \ri\cG)^{\T^n}$ inherits a Poisson bracket
from  $(C^\infty(\cH^o \times \ri\cG), \{\ ,\ \}_\Li)$. Specifically, the induced Poisson bracket
of two smooth gauge invariant functions $F,H\in C^\infty(\cA^o \times \ri\cG)^{\T^n}$
can be determined by the standard `extend--compute--restrict' algorithm.
That is, one first extends $F$ and $H$ arbitrarily from the first class constraint surface, then
determines the Poisson bracket of the extended functions, $F^\ext$ and $H^\ext$, and finally restricts the
result to the $\phi=0$ constraint surface. This gives a well-defined Poisson bracket.
For the theory of Hamiltonian reduction, we recommend the books \cite{HT,OR}.

Our situation is extremely simple, since by decomposing any $w\in\cH^o$ as
\be
w = \pi_\cA(w) + \pi_\cT(w)
\ee
we can naturally extend any $F\in C^\infty(\cA^o\times \ri\cG)^{\T^n}$ to the
phase space $\cH^o\times \ri \cG$  by declaring that
\be
F^\ext(w,L):= F(\pi_\cA(w), L).
\label{Fext}\ee
Here and below, the various  projection operators $\pi_{\cA}$, $\pi_{\cT}$ etc.~rely
on the decompositions (\ref{I5}), (\ref{I6}).
The function $F^\ext$ defined in this manner belongs to
$C^\infty(\cH^o\times \ri\cG)^{\T^n}$.
It follows immediately from (\ref{LiPB}) that the induced Poisson bracket
\be
\{ F,H\}_2(q,L):= \{ F^\ext, H^\ext\}_\Li(q,L), \qquad \forall (q,L)\in \cA^o \times \ri\cG,
\label{FextPB}\ee
is given by the formula displayed in Theorem 1.
A further consequence of the reduction is that we have
\be
V_m[F] = \{F, H_m\}_2,
\ee
where $H_m$ is now regarded as a gauge invariant function on the $\phi=0$ constraint surface.
Since $V_m[H_\ell]=0$,  $\{ H_\ell, H_m\}_2 =0$ results as well.

In conclusion, by following \cite{Li2}, we have explained the part of the
statements of Theorem 1 pertaining to the Poisson bracket $\{\ ,\ \}_2$.

\section{The Braden--Hone system as a spin Sutherland model}

The invariant bilinear form $\langle\ ,\ \rangle_\R$ on $\cG^\C_\R$ can be used to define a bi-invariant
semi-Riemannian metric on the group manifold $G_\R^\C$, whose geodesics
are the orbits of the one-parameter subgroups of $G_\R^\C$ with respect to right-multiplication
(or, equivalently, left-multiplication).
Hamiltonian reductions of such `free geodesic motion' giving rise to spin Sutherland models
have been investigated previously, for example in \cite{FP2}.
Building on this reference, we here explain  how the Braden--Hone system results from reduction.

We are going to reduce the phase space $T^* G_\R^\C \times \cG^*$,
where $\cG^*$ is added for technical convenience (akin to the so-called
shifting trick of symplectic reduction \cite{OR}).
We trivialize the cotangent bundle by left-translations, and identify  $\cG^\C_\R$ and $\cG$
with their own dual spaces
by means of the form $\langle\ ,\ \rangle_\R$.
Thus our unreduced phase space, $P$, is
\be
P:= T^* G^\C_\R \times \cG^* \equiv G^\C_\R \times \cG_\R^\C \times \cG = \{ (g,J, \xi)\}
\label{P}\ee
endowed with its standard Poisson structure. For any  smooth real functions
$f$ and $h$ on $P$, the Poisson bracket reads
\be
\{f,h\}_P(g,J,\xi) = \langle D_g' f, \nabla_J h\rangle_\R -\langle D_g' h, \nabla_J f\rangle_\R
-\langle J, [\nabla_J f, \nabla_J h]\rangle_\R + \langle \xi, [\nabla_\xi f, \nabla_\xi h]\rangle_\R,
\label{PPB}\ee
where $\nabla_J f$ and $\nabla_\xi f$ are $\cG_\R^\C$-valued and $\cG$-valued
`partial gradients' defined by
 using $\langle\ ,\ \rangle_\R$, and $D_g'f(g,J,\xi)\in \cG_\R^\C$ is defined by
\be
 \langle X, D_g'f(g,J,\xi)\rangle_\R :=\dt f(g e^{tX}, J, \xi), \qquad \forall X\in \cG^\C_\R.
\ee

We single out the `free Hamiltonians' $h_m$,
\be
h_m(g,J,\xi):= \frac{1}{m}\Re\tr(J^m),\qquad \forall m\in \N,
\label{hm}\ee
which form an Abelian algebra under the Poisson bracket.
Denote by $\V_m$ the Hamiltonian vector field generated by $h_m$. It has the explicit form
\be
\V_m[g] = g J^{m-1},\quad \V_m[J]=0,\quad \V_m[\xi]=0.
\label{35}\ee
Here, $\V_m[g]$ etc.~are understood as derivatives of evaluation functions.
This means that $\V_m[g]$ collects the derivatives of  (the real and imaginary parts of) the
matrix elements of $g$, which are regarded as functions on $P$
 (see also footnote 1).

We reduce relying on the action of the symmetry group $G\times G$ on $P$,
where $G:= \UN(n)$.
 We let
$(\eta_L, \eta_R)\in G\times G$ act on $P$ by the diffeomorphism $\Psi_{\eta_L, \eta_R}$,
\be
\Psi_{\eta_L, \eta_R}(g,J,\xi) := (\eta_L g \eta_R^{-1}, \eta_R J \eta_R^{-1}, \eta_L \xi \eta_L^{-1}).
\ee
This is a Hamiltonian action. The corresponding moment map
$\Phi: P \to \cG \oplus \cG$ is given by
\be
\Phi(g,J,\xi) = \left( \pi_{\cG}( g J g^{-1})+ \xi, -\pi_{\cG}(J)\right).
\label{Phi}\ee
We impose the constraint $\Phi=0$, and then divide by the `gauge transformations'
associated with $G\times G$. The gauge invariant functions on $P_0:= \Phi^{-1}(0)$
inherit a Poisson structure, and the vector fields $\V_m$ induce commuting derivations
of $C^\infty(P_0)^{G\times G}$.

It is worth noting that the `partial moment map constraint' $\pi_\cG(J)=0$
enforces the reduction of $T^* G_\R^\C$ to $T^*(G_\R^\C/G)$, which underlies the link
to the approach of the paper \cite{FP2}.
Indeed, one could perform the reduction by $G\times G$ in two steps, and first imposing
only $\pi_\cG(J)=0$ would lead, in effect, to
the starting point of the reduction studied in \cite{FP2}.

From now on we restrict our attention to the dense open submanifold $P^\reg\subset P$, which is
characterized by the condition that $g$ can be decomposed as
\be
g= \eta_L^{-1} e^{q} \eta_R
\quad \hbox{with}\quad q\in \cA^o,\, \eta_L, \eta_R\in G.
\label{gdec}\ee
In this decomposition $q$ is unique, while the pair $(\eta_L, \eta_R)$ is
unique up to the ambiguity of its possible replacement by $(\eta \eta_L, \eta \eta_R)$
 with an arbitrary $\eta\in \T^n$.
It is plain that every gauge orbit lying in $P^\reg_0$ has representatives in the following `gauge slice'
$S\subset P_0^\reg$:
\be
S:= \{ (e^q, \cL, \xi)\in P_0 \mid q\in \cA^o,\,\, \cL \in \ri \cG\}.
\label{S}\ee
On the elements of $S$, the condition  $\Phi=0$ (\ref{Phi}) translates into the equation
\be
\xi + (\sinh\ad_q)( \cL)=0,
\label{res}\ee
and we have a residual gauge action of $\T^n$ on $S$, given by the maps $\Psi_{\eta,\eta}$:
\be
\Psi_{\eta,\eta}( e^{q}, \cL, \xi) = (e^q, \eta \cL \eta^{-1}, \eta \xi \eta^{-1}),
\qquad \eta \in \T^n.
\ee
Moreover, we see from the constraint equation (\ref{res}) that
\be
\pi_\cT(\xi)=0 \quad\hbox{and} \quad \pi_{\cT^\perp}(\xi) = - (\sinh\ad_q) (\pi_{\cA^\perp}(\cL)),
\ee
which provides a parametrization of $S$ by the `free variables' $(q,\cL)\in \cA^o \times \ri\cG$.
An important consequence is the chain of identifications
\be
C^\infty(P_0^\reg)^{G\times G} \Longleftrightarrow C^\infty(S)^{\T^n}
\Longleftrightarrow C^\infty(\cA^o \times \ri \cG)^{\T^n}.
\label{id1}\ee

We note in passing that the linear operator $(\sinh\ad_q)$ is zero on $(\cA+ \cT)$,
while it maps $\cT^\perp$ to $\cA^\perp$ and $\cA^\perp$ to $\cT^\perp$ in an
invertible manner, due to the regularity of $q$ (see (\ref{I3}) and (\ref{I6})).
When below we write $(\sinh \ad_q)^{-1}$,
then we mean the unique inverses of these restricted operators.

Now we explain how the commuting vector fields $\V_m$ (\ref{35}) descend
to the Braden--Hone hierarchy (\ref{Vm}).
The vector fields $\V_m$ are tangent to $P_0$, but are not tangent to $S$.
However, since we are interested in the evolution of the gauge invariant `observables',
we can cure this non-tangency by adding a suitable infinitesimal gauge transformation
to $\V_m$.  The latter is given by a pair of $\cG$-valued
functions $Y^L$ and $Y^R$ on $S$, which are required to ensure that the
following vector field $\V_m^S$ is tangent to $S$:
\be
\V_m^S[e^{q}] = e^q\cL^{m-1} + Y^Le^q - e^q Y^R,
\quad
\V_m^S[\cL] = [Y^R, \cL].
\label{314}\ee
The first equation determines the pair $(Y^L, Y^R)$ up to shifts defined by adding $(Y,Y)$,
where $Y$ is an arbitrary $\cT$-valued function on $S$. This ambiguity corresponds to the
residual gauge transformations acting on $S$.
Indeed, the expression
\be
e^{-q} \V_m[e^q] = \cL^{m-1} - (\sinh \ad_q)(Y^L) + (\cosh \ad_q)(Y^L) - Y^R
\ee
must belong to $\cA$, and this condition has the following solution:
\be
 Y^L= (\sinh \ad_q)^{-1} (\pi_{\cA^\perp}(\cL^{m-1})), \qquad
Y^R = \cR(q)(\cL^{m-1}),
\label{316}\ee
with $\cR(q)$ defined in (\ref{R}).
We here used that $\cL^{m-1}$ is Hermitian, i.e., $\cL^{m-1}\in \ri \cG$.
The reduced dynamics is obtained by substitution of (\ref{316}) into (\ref{314}).
The next proposition summarizes the outcome of our line of reasoning.

\medskip\noindent
{\bf Proposition 4.}
\emph{The derivative of the gauge invariant function
$F\in C^\infty(\cA^o \times \ri \cG)^{\T^n}$
 with respect to the reduction
of the Hamiltonian vector field $\V_m$ (\ref{35}) is encoded by the formula
\be
\V^S_m[F] = \langle \V^S_m[q], \nabla_1 F \rangle_\R + \langle \V_m^S[\cL], \nabla_2 F\rangle_\R,
\ee
where
\be
\V^S_m[q] = \pi_{\cA} (\cL^{m-1})
\quad\hbox{ and}\quad
\V^S_m[\cL] = [ \cR(q)(\cL^{m-1}), \cL].
\label{VS}\ee
Comparison with equation (\ref{Vm}) shows that the reduction yields the Braden--Hone hierarchy
defined in the Introduction. Specifically, the vector field
$\V_{m+1}^S$ reproduces $V_m$ (\ref{Vm}).}
\medskip

The gauge slice $S$ (\ref{S}) has two distinguished parametrizations. The first one is by the variables
$(q,\cL)\in \cA^o \times \ri\cG$, and the second one is by the variables
\be
(q,p,\xi_\perp) \in \cA^o  \times \cA \times \cT^\perp,
\ee
which are related to $(q,\cL)$ by the equation
\be
\cL = p - (\sinh\ad_q)^{-1}(\xi_\perp).
\label{cL1}\ee
The correspondence between $(q,\cL)$ and $(q,p, \xi_\perp)$ is a $\T^n$-equivariant
diffeomorphism
between $\cA^o \times \ri \cG$ and $\cA^o \times \cA \times \cT^\perp$.
Of course, $q$ and $p$ are $\T^n$-invariants, while $\xi_{\perp}$ transforms by
conjugation.
This may be used to extend the chain of identifications (\ref{id1}) as
\be
C^\infty(P_0^\reg)^{G\times G} \Longleftrightarrow C^\infty(S)^{\T^n}
\Longleftrightarrow C^\infty(\cA^o \times \ri \cG)^{\T^n}
 \Longleftrightarrow C^\infty(\cA^o \times \cA \times \cT^\perp)^{\T^n}.
\label{id2}\ee
By identifying  $\cA^*$ with $\cA$ using $\langle\ ,\ \rangle_\R$,
 $\cA^o \times \cA$ can be taken as a model of  $\T^*\cA^o$,  which
carries a symplectic form. Moreover, $C^\infty(\cT^\perp)^{\T^n}$ is a Poisson algebra,
 equipped with the reduction of the Lie--Poisson bracket of $\cG^*\equiv \cG$ defined by
 the first class constraint $\pi_\cT(\xi)=0$.

\medskip \noindent
{\bf Proposition 5.}
\emph{Let us parametrize $S$ (\ref{S}) by the variables $q,p,\xi_{\perp}$ using (\ref{cL1}) and consider
two gauge invariant functions $F,H\in C^\infty(\cA^o \times \cA \times \cT^\perp)^{\T^n}$.
In terms of these functions, the reduced Poisson bracket arising from the Poisson structure (\ref{PPB}) on
$P$ can be written as
\be
\{ F, H\}_P^\red = \langle \nabla_q F, \nabla_p H \rangle_\R - \langle \nabla_q H, \nabla_p F \rangle_\R
+ \langle \xi_\perp, [\nabla_\xi F, \nabla_\xi H]\rangle_\R.
\label{SPB1}\ee
Here, $\nabla_qF$, $\nabla_p F$ are the obvious $\cA$-valued gradients, and
$\nabla_\xi F$ can be taken from $\cT^\perp\subset \cG$, applying the definition
\be
\langle X, \nabla_\xi F(q,p,\xi_\perp)\rangle_\R  = \dt F(q,p,\xi_\perp + t X),
\qquad\forall X\in \cT^\perp.
\ee}

 The statement of the proposition  is basically a special case of more general results proved in \cite{FP2}.
In \cite{FP2} analogous reductions were studied, but restricting $\xi$
to an arbitrary coadjoint orbit $\cO$ of $G$ from the very beginning. A
counterpart of the formula (\ref{SPB1}) can be obtained by evaluation
of the restriction of the symplectic form of  $T^* G_\R^\C \times \cO$ to the gauge slice where
$g = e^{q}$, $q\in \cA^o$.  This proves the claim in our case, too, since
$\cO \subset \cG^*$ is a symplectic leaf.

We  stress that in the formula (\ref{SPB1}) one can also determine
$\nabla_\xi F$ utilizing an arbitrary extension of $F$ from $T^*\cA^o \times \cT^\perp$
to $T^*\cA^o \times \cG$,
computing the $\cG$-valued derivative $\nabla_\xi$ there, and restricting the result.
This leads to an ambiguity regarding the $\cT$-components of
the derivatives with respect to $\xi$, which eventually drops out
on account of the $\T^n$-invariance of $F$ and $H$ on $T^*\cA^o\times \cT^\perp$.

Finally, we turn to the description of the reduced Poisson bracket in terms of
gauge invariant functions $F,H \in C^\infty(\cA^o \times \ri \cG)^{\T^n}$.
For this purpose, we need an auxiliary lemma.

\medskip \noindent
{\bf Lemma 6.}  \emph{Define the map $\cL: \cA^o \times \cA \times \cG \to \ri \cG$ by
 extension of the formula (\ref{cL1}), i.e., by
\be
\cL(q,p,\xi):= p - (\sinh\ad_q)^{-1}(\pi_{\cT^\perp}(\xi)).
\label{cL2}\ee
For any $X\in \cA$, $Y\in \cG$ and  $Z\in \ri \cG$, define $q^X:= \langle X, q\rangle_\R$,
$\xi^Y:= \langle Y, \xi \rangle_\R$ and $\cL^Z:= \langle Z, \cL\rangle_\R$.
Regarding these as functions on the Poisson manifold $T^* \cA^o \times \cG^*$,
the following formulae hold:
\be
\{ q^X, \cL^Z\} = \langle X, Z \rangle_\R, \quad  \{\cL^Z, \xi^T\} = \cL^{[T,Z]}
\quad \hbox{for all}\quad X\in \cA,\,T\in \cT,\,Z\in \ri \cG,
\label{prePB}\ee
 and
\be
\{ \cL^{Z_1}, \cL^{Z_2}\} = \cL^{[Z_1,Z_2]_{\cR(q)}} +
\langle \pi_\cT(\xi), [\cW(\ad_q)\pi_{\cA^\perp}(Z_1), \cW(\ad_q)\pi_{\cA^\perp}(Z_2)]\rangle_\R
\label{rPB}\ee
for all $Z_1, Z_2\in \ri \cG$, where $\cW(\ad_q): \cA^\perp \to \cT^\perp$ is
given by $\cW(\ad_q)Z:= (\sinh\ad_q)^{-1}(Z)$.
The notation $[Z_1, Z_2]_{\cR(q)}$ is defined after (\ref{I14}).  Of course, the Poisson brackets
between any  functions of $q$ and $\xi_\cT=\pi_\cT(\xi)$ are zero.}

\medskip
The formulae (\ref{prePB}) are easy consequences of the parametrization of $\cL$ (\ref{cL2}),
using the canonical Poisson brackets between the components of $q$, $p$ and the Lie--Poisson
bracket
\be
\{\xi^{Y_1}, \xi^{Y_2}\} = \xi^{[Y_1,Y_2]}, \qquad  \forall Y_1, Y_2\in \cG.
\label{trivi}\ee
The verification of equation (\ref{rPB})  is straightforward, but rather tedious\footnote{
This calculation was performed by B.G. Pusztai in a more general case during our collaboration in 2006.} .
This is omitted to save place.  Note that
an analogous result was given in  \cite{FP2}, and more recently
also in \cite{KLOZ}.

\medskip \noindent
{\bf Remark 7.}
The formulae (\ref{prePB}) and (\ref{rPB}) can be viewed as the defining relations of a Poisson structure on the manifold
\be
\cA^o \times \ri\cG \times \cT = \{ (q,\cL, \xi_\cT)\}.
\ee
This is nothing but the Poisson structure of $T^* \cA^o \times \cG^*$ transferred to
$\cA^o \times \ri \cG \times \cT$ by means of the invertible change of variables
$(q, p, \xi) \leftrightarrow (q,\cL, \xi_\cT)$ (\ref{cL2}).
Then  the Poisson bracket on $C^\infty(\cA^o \times \ri \cG)^{\T^n}$ can be represented
as the reduction of $(C^\infty(\cA^o \times \ri\cG \times \cT), \{\ ,\ \})$  defined by the first class constraint $\xi_\cT=0$.
Indeed, this follows from Proposition 5.
 The identifications (\ref{id2}) give rise to
 alternative descriptions of the Poisson bracket on $C^\infty(P_0^\reg)^{G\times G}$, which descends
 from $\{\ ,\ \}_P$ (\ref{PPB}).

\medskip\noindent
{\bf Proposition 8.}
\emph{Let us parametrize $S$ (\ref{S}) by the variables $q,\cL$ and consider
two gauge invariant functions $F,H\in C^\infty(\cA^o \times \ri \cG)^{\T^n}$.
In terms of these functions, the reduced Poisson bracket descending from the Poisson structure (\ref{PPB}) on
$P$ can be written as
\be
\{ F, H\}_P^\red =  \langle  \nabla_1 F,  \nabla_2 H \rangle_\R
- \langle  \nabla_1 H,  \nabla_2 F\rangle_\R
+\langle \cL, [ \nabla_2 F,  \nabla_2 H]_{\cR(q)}   \rangle_\R,
\label{redPB2}\ee
which coincides with the Poisson bracket $\{\ ,\ \}_1$ given by equation (\ref{I14}) of Theorem 1.
Denoting the restriction of $h_m$ (\ref{hm}) to $S$ by $H_m$, we obtain the following consequence
of the reduction:
\be
\V_{m}^S[F] = \{ F, H_m  \}_P^\red,
\label{redHam}\ee
 which implies the second equality in (\ref{I15}), since $V_m = \V_{m+1}^S$ by (\ref{VS}).}

\begin{proof}
According to Remark 7,
the Poisson bracket on $C^\infty(\cA^o \times \ri \cG)^{\T^n}$ can be
calculated as follows.  Regard $q$, $\cL$ and $\xi_\cT$ as independent variables, determine the Poisson brackets
of the functions
of $q$ and $\cL$ by utilizing the formulae of Lemma 6, and impose the constraint $\xi_\cT=0$ at the end of the calculation.
This algorithm proves the claim (\ref{redPB2}).
The equality (\ref{redHam}) is a consequence of the theory of Hamiltonian reduction:
$\V_m^S$ represents the reduced
Hamiltonian vector field descending from $\V_m$ (\ref{35}) and $H_m$ is the corresponding reduced
Hamiltonian (note that $\Re\tr(\cL^m)= \tr (\cL^m)$ since $\cL$ is Hermitian).
\end{proof}

\section{Proof of Theorem 2}

We begin the proof of Theorem 2 by recapitulating the core points of the preparations.

In equation (\ref{LiPB}) we have introduced the Poisson manifold
\be
(\cH^o \times \ri \cG, \{\ ,\ \}_\Li).
\label{P1}\ee
The coordinate functions on this manifold can be taken to be the components $q_i$ of $q=\Re(w)$,
the components of $\Im(w)$, and the functions
$L_a:= \langle Z_a, L \rangle_\R$ associated with a basis $\{ Z_a\}$ of $\ri \cG$.
The Poisson algebra
\be
(C^\infty(\cA^o \times \ri\cG)^{\T^n}, \{\ ,\ \}_2)
\label{P2}\ee
results from (\ref{P1}) by imposing the first class constraint $\Im(w)=0$, which is equivalent to
the equality $w=q\in \cA^o$.
The reduced Poisson algebra is completely determined by the Poisson brackets between such gauge
invariant functions  that depend only on $L$ or only on $q$.

Let us extend the derivation $\Der$ (\ref{Der}) to a derivation of $C^\infty(\cH^o \times \ri\cG)$
by declaring that
 the derivatives of all components
of $\Im(w)$ are zero.
Then introduce the bracket $\{F,H\}_\Li^{\Der}$ by
\be
\{F,H\}^{\Der}_\Li := \Der[\{ F,H\}_\Li] - \{ \Der[F], H\}_\Li - \{ F, \Der[H]\}_\Li,
\label{P3}\ee
and similarly introduce $\{F,H\}^{\Der}_2 := \Der[\{F,H\}_2] - \{ \Der[F], H\}_2 - \{F,\Der[H]\}_2$.
These brackets automatically satisfy the Leibniz and the anti-symmetry properties,
but the Jacobi identity is not guaranteed.

Focusing on  gauge invariant functions $\cP_1$ and $\cP_2$ depending only on $L$, we have
\be
\{\cP_1, \cP_2\}_2(q,L) =\{\cP_1, \cP_2\}_\Li(q,L)= \sum_{a,b}  \{L_a, L_b\}_\Li(q,L)
\left(\frac{\pa \cP_1}{\pa L_a}\frac{\pa \cP_2}{\pa L_b}\right)(L).
\label{P4}\ee
This is a special case of the formula (\ref{FextPB}).  We employ the trivial extension (\ref{Fext}),
and thus we do not need to
use a separate notation for the extended functions. For example, a $\T^n$-invariant polynomial formed
out of the components of $L$ can be regarded both as a function on $\cA^o \times \ri \cG$ and as a function on
$\cH^o \times \ri\cG$.
As a consequence of the formula  (\ref{P4}) and the definition of $\{\ ,\ \}_2^{\Der}$, we also have
\be
\{ \cP_1,\cP_2\}_2^{\Der}(q,L) = \sum_{a,b}
\{L_a, L_b\}_\Li^{\Der}(q,L) \left(\frac{\pa \cP_1}{\pa L_a}\frac{\pa \cP_2}{\pa L_b}\right)(L).
\label{P5}\ee
This shows that all information about $\{ \cP_1,\cP_2\}_2^{\Der}$ is contained in $\{L_a, L_b\}_\Li^{\Der}$.

On the other hand, as was proved in Section 3, the Poisson algebra
\be
(C^\infty(\cA^o\times \ri \cG)^{\T^n},\{\ ,\ \}_1)
\label{P6}\ee
is a reduction of the Poisson algebra
\be
(C^\infty(\cA^o\times \ri\cG \times \cT), \{\ ,\ \}),
\ee
where $\{\ ,\ \}$  denotes the Poisson bracket given by Lemma 6 (see also Remark 7).
The reduction is defined by the first class constraint $\xi_\cT=0$.
Accordingly, we have
\be
\{\cP_1, \cP_2\}_1(q,L) = \sum_{a,b}  \{L_a, L_b\}(q,L,\xi_\cT=0)
\left(\frac{\pa \cP_1}{\pa L_a} \frac{\pa \cP_2}{\pa L_b}\right)(L).
\label{P7}\ee

Now, Theorem 2 claims that
\be
 \{ \cP_1, \cP_2\}_1(q,L)= \{\cP_1, \cP_2\}_2^{\Der} (q,L),
\label{P8}\ee
and we see by comparison of (\ref{P5}) and (\ref{P7})  that this follows if we can verify that
\be
\{ L_a, L_b\}(q,L,\xi_\cT=0) = \{ L_a, L_b\}_\Li^{\Der}(q,L).
\label{P9}\ee
Since $L_a \equiv L^{Z_a}$,  Lemma 6 gives
\be
\{ L_a, L_b\}(q,L,\xi_\cT=0) = \langle L, [\cR(q) Z_a, Z_b] + [Z_a, \cR(q) Z_b] \rangle_\R.
\label{P10}\ee
We compute the right-hand side of (\ref{P9}) from the definition
(\ref{LiPB}) noting that by (\ref{Der})  $\Der[L]=\cD[L_a Z^a]:= \cD[L_a] Z^a=\1_n$ is the unit matrix
(the basis $\{ Z^a\}$ is dual to $\{Z_a\}$).
We find
\bea
&& \{L_a, L_b\}_\Li^\cD(q,L) = - 2 \langle \cR(q)(LZ_a), Z_b \rangle_\R -
2\langle  \cR(q) Z_a, L Z_b\rangle_\R \nonumber\\
 && \phantom{  \{L_a, L_b\}_\Li^\cD(q,L)}
 = 2 \langle L, Z_a \cR(q) Z_b  - Z_b \cR(q) Z_a \rangle_\R.
\eea
Using that $(Z_a \cR(q) Z_b)^\dagger = - (\cR(q) Z_b) Z_a$, we get
\be
\langle L,  Z_a\cR(q)Z_b \rangle_\R  = - \langle L, (\cR(q) Z_b) Z_a\rangle_\R.
\ee
 In this way we confirm (\ref{P9}), and thus
the claim (\ref{P8}) holds.

 To finish the proof of the claim (\ref{claim1}), it is sufficient to verify
 the equalities
 \be
  \{ \cF, \cP\}_1(q,L)=\{ \cF, \cP\}_2^{\Der}(q,L)
 \quad\hbox{and}\quad
 \{\cF_1, \cF_2\}_1(q,L)=\{ \cF_1, \cF_2\}^{\Der}_2(q,L)
 \label{P11}\ee
 for invariant functions $\cP$ of $L$ and arbitrary smooth functions
 $\cF$, $\cF_i$ depending only on $q$.
 These verifications are in principle similar to the above, but are computationally  simpler.

 Turning to the claim (\ref{claim2}), now we extend the derivation
 $\cD$ (\ref{Der}) to $C^\infty(\cA^o \times \ri \cG \times \cT)$ by setting $\cD[\xi_\cT]:= 0$,
 and then
 define $\{\ ,\ \}^{\Der}$ analogously to (\ref{P3}).
 With $Q:= \sum_{i=1}^n q_i$, we notice from Lemma 6 that
 \be
 \cD[F] = \{ Q , F\},
 \qquad
 \forall F\in  C^\infty(\cA^o \times \ri \cG \times \cT).
 \ee
 This implies that  $ \{F,H\}^\cD =0$ for all functions.
 From this, referring to Remark 7 and Proposition 8, the  claim (\ref{claim2}) follows.

Since the compatibility of $\{\ ,\ \}_2$ and
$\{\ ,\ \}_1=\{\ ,\ \}_2^{\Der}$ is a consequence of  Lemma 3 given in
the Introduction, the proof of Theorem 2 is now complete.

\section{Conclusion}

In this paper we combined two approaches to the Braden--Hone hierarchy (\ref{I8}), and have shown that together they endow
these evolution equations with a bi-Hamiltonian structure.

The approach based on Hamiltonian reduction of free motion on $G_\R^\C$ enjoys the attractive
feature that the  initial free flows are complete,
and this  is automatically inherited by the reduced flows. However, to realize the completeness one might
need to drop the restriction to regular elements in the decomposition (\ref{gdec}), which is valid
only on a dense open submanifold. This issue requires further investigation.
We note only that the free flow generated by $h_m$ (\ref{hm}) reads
\be
g(t) = g(0)\exp(t J^{m-1}),\quad J(t) =J(0), \quad \xi(t)=\xi(0),
\ee
 from which the flows of the Braden--Hone system result by the standard projection method.

The hyperbolic Braden--Hone hierarchy that we have studied admits a trigonometric version,
for which similar results are expected to hold. This, and the question of generalizations
to other Lie algebras and twisted cases, will be investigated elsewhere. Previous works
relevant to such a future study include \cite{F,FP1,FP2,Li2}.

Another very interesting unexplored  aspect concerns the
integrability (both in Liouville and in non-commutative sense)
of the bi-Hamiltonian Braden--Hone hierarchy.
In this respect,
since the notions of integrability are usually formulated on symplectic manifolds,
one should
investigate
the symplectic leaves of the alternative Poisson structures
on $(\cA^o \times \ri \cG)/\T^n$.
It is also natural to ask what happens to the Poisson bracket $\{\ ,\ \}_2$ if one
restricts to a symplectic leaf of $\{\ ,\ \}_1$ (and vice versa)?
For the investigation of non-commutative integrability,
one may adapt the approach of the papers \cite{Res1,Res},   where  non-commutative (degenerate)
integrability was proven for a family of trigonometric spin Calogero--Moser--Sutherland systems.
It is known (see e.g.~\cite{FK}) that restriction to a minimal coadjoint orbit
$\cO \subset \cG^*$ in (\ref{P}) leads
to the spinless hyperbolic Sutherland model, with its standard Poisson structure
arising from $\{\ ,\ \}_1$.
There should be a way to recover the spinless hyperbolic Ruijsenaars--Schneider model \cite{RS}
via restriction to a small symplectic leaf of the structure $\{\ ,\ \}_2$.
When studying all these questions, one should of course take note of the fact
that $\ri \cG/\T^n$  is not a smooth manifold, but is a union of smooth
strata, since the $\T^n$-action has several different orbit types \cite{OR}.
One should apply
the theory of singular Hamiltonian reduction \cite{OR,SL, Sn} to
 uncover the global structure of the reduced system that emerges
from the geodesic motion on $G_\R^\C$.

Finally, it is an open problem if there is any relation between the
results  of this paper and the previous works \cite{Bar,FaMe} devoted to the bi-Hamiltonian structure
of the (spinless) rational Calogero--Moser system.

\bigskip
\begin{acknowledgements} I wish to thank L.-C.~Li for correspondence
on related matters, which aroused my interest in the bi-Hamiltonian
issue.
I am grateful to J.~Balog, I.~Marshall and B.G.~Pusztai for remarks on the manuscript.
 This work was supported  by the
Hungarian Scientific Research Fund (OTKA) under the grant
 K-111697.
\end{acknowledgements}

\appendix
\section{Commuting derivations of gauge invariants}
\label{sec:A}

In this appendix we show by a direct method that the Braden--Hone hierarchy (\ref{I8}) induces
commuting derivations of the gauge invariant functions forming $C^\infty(\cA^o \times \ri \cG)^{\T^n}$.

We start by noting that $\cR: \cH^o \to \operatorname{End}(\cG_\R^\C)$,
defined by (\ref{R}),
is anti-symmetric with respect to the bilinear form (\ref{I4}) and is $\cH$-invariant (see (\ref{cH}))
in the sense that
\be
[\ad_T, \cR(w)] = 0,
\quad \forall T\in \cH, \quad w \in \cH^o.
\ee
Let us consider the derivation $V_m$ of $C^\infty(\cA^o \times \ri \cG)$ specified by the rules (\ref{Vm}).
The derivative $V_m[F]$ of $F\in C^\infty(\cA^o \times \ri\cG)$ reads
\be
V_m[F] = \langle V_m[q], \nabla_1 F\rangle_\R + \langle V_m[L], \nabla_2 F\rangle_\R.
\ee
Using the invariance property of $\cR$, it is easily shown that $V_m$ preserves the
ring of gauge invariant functions, $C^\infty(\cA^o\times \ri \cG)^{\T^n}$.

\medskip\noindent
{\bf Proposition A1.}
\emph{The commutator of two derivations $V_m$ and $V_\ell$ satisfies
\be
(V_m \circ V_\ell - V_\ell \circ V_m)[q]=0,
\qquad
(V_m \circ V_\ell - V_\ell \circ V_m)[L] = [ T_{m,\ell}(q,L), L]
\label{A}\ee
with a certain function $T_{m,\ell}(q,L) \in \cT$,
representing an infinitesimal $\T^n$ gauge transformation.
Consequently, the restrictions of the derivations $V_m$ and $V_\ell$ to
$C^\infty(\cA^o\times \ri \cG)^{\T^n}$ commute with each other, for any $m,\ell \in \N$.
}

\medskip\noindent
\begin{proof}
We obtain from the definitions
\be
(V_m \circ V_\ell - V_\ell \circ V_m)[q]=\left([ \cR(q)L^m, L^\ell] - [\cR(q)L^\ell, L^m]\right)_\cA,
\ee
where the subscript $\cA$ refers to the decomposition (\ref{I6}).
Taking any $A\in \cA$, by using the invariance and anti-symmetry of $\cR$, we can write
\be
\langle A, [ \cR(q)L^m, L^\ell]\rangle_\R = \langle [A, \cR(q) L^m], L^\ell\rangle  =
\langle \cR(q)[A,L^m], L^\ell \rangle_\R = - \langle A, [L^m, \cR(q)L^\ell]\rangle.
\ee
From this, we get
\be
\langle A, (V_m \circ V_\ell - V_\ell \circ V_m)[q]\rangle_\R=0,\qquad \forall A\in \cA,
\ee
which is equivalent to the first equality in (\ref{A}).

Next, a simple calculation gives
\be
(V_m \circ V_\ell - V_\ell \circ V_m)[L]= [T_{m,\ell}, L]
\ee
with
\be
T_{m,\ell} =
\cR([\cR L^m, L^\ell] + [L^m, \cR L^\ell]) - [\cR L^m,\cR L^\ell]
+ (\nabla_{(L^m)_\cA} \cR) L^\ell -  (\nabla_{(L^\ell)_\cA}\cR) L^m.
\label{A9}\ee
We simplified the notation by omitting the argument $q$ of $\cR$.
For any $T\in \cH$,
$\nabla_T \cR$ denotes the directional derivative of $\cR$.
In order to show that $T_{m,\ell}$ (\ref{A9}) belongs to $\cT$, we recall \cite{FP1} that
the modified classical dynamical Yang--Baxter equation, satisfied by $\cR: \cH^o \to \operatorname{End}(\cG_\R^\C)$,
can be written as follows:
\be
 \cR([X,\cR Y] + [\cR X,Y]) - [\cR X, \cR Y]  +   (\nabla_{X_\cH}\cR) Y - (\nabla_{Y_\cH} \cR) X =  [X,Y] +
\langle X, (\nabla \cR) Y\rangle,
\label{CDYBE}\ee
for all $ X,Y\in \cG_\R^\C$.
Using a basis $A_i$ of $\cA$ and a basis $T_a$ of $\cT$, with corresponding dual bases
$A^i$ and $T^a$ with respect to the bilinear form (\ref{I4}), we have
\be
\langle X, (\nabla \cR) Y\rangle := \sum_i A^i \langle X, (\nabla_{A_i}\cR) Y\rangle
+ \sum_a T^a \langle X, (\nabla_{T_a} \cR) Y\rangle.
\label{A11}\ee
To determine $T_{m,\ell}$ (\ref{A9}) from (\ref{CDYBE}), we have to evaluate the expression (\ref{A11})
for $X=L^m$ and $Y= L^\ell$, at $q\in \cA^o$.
Now, for any $T\in (\cA + \cT)$, we find that $(\nabla_T \cR)(q)$ acts non-trivially on
 $(\cA^\perp + \cT^\perp)$
by the operator $\ad_T \circ f(\ad_q)$,
where $f$ is the analytic function
\be
f(z) = \frac{d\coth(z)}{dz} = - \frac{1}{\sinh^2(z)}.
\ee
This implies that
\be
\langle L^m, (\nabla \cR)(q) L^\ell\rangle = \sum_a T^a \langle L^m, (\nabla_{T_a} \cR)(q) L^\ell\rangle_\R
= [f(\ad_q) ((L^\ell)_{\cA^\perp}), L^m]_\cT.
\label{A14}\ee
The terms $\langle L^m,(\nabla_{A_i}\cR)(q) L^\ell\rangle_\R$ of (\ref{A11})
vanish, because $L^m\in \ri \cG$ and $(\nabla_{A_i}\cR)(q)L^\ell \in \cG$.
\end{proof}

\section{Rewriting the Poisson bracket formula of L.-C. Li}

In this appendix we recall the Poisson bracket given by the formula (5.8) in \cite{Li2},
and explain its relation to our formula (\ref{LiPB}).

Consider the manifold
\be
\cM:=\cH^o \times G_\R^\C \times \cH^o =\{ (u,g,v) \}
\ee
and its submanifold
\be
\cM_{\herm}:= \{ (u,L, u^\dagger)\mid u\in \cH^o, \, L\in \fH\},
\ee
where $\fH$ is the subset of the Hermitian elements of $G^\C_\R= \GL(n,\C)$.
Take any real function $f\in C^\infty(\cM_{\herm})$, and extend it arbitrarily to
an element $f^\ext\in C^\infty(\cM)$.
Define the $\cH$-valued derivatives $\delta_i f^\ext$ $(i=1,2)$ and the $\cG^\C_\R$-valued derivatives
$Df^\ext$ and $D' f^\ext$ by the requirements
\be
\langle \xi, \delta_1 f^\ext(u,g,v)\rangle_\R +
\langle \eta, \delta_2 f^\ext(u,g,v)\rangle_\R:=
\dt f^\ext(u + t \xi, g, v + t \eta ), \quad \forall \xi, \eta\in \cH,
\ee
and
\be
\langle X, Df^\ext(u,g,v) \rangle_\R + \langle Y, D'f^\ext(u,g,v)\rangle_\R
:=\dt f^\ext(u, e^{tX} g e^{tY}, v), \quad \forall X,Y\in \cG_\R^\C.
\ee
Then, following \cite{Li2},  introduce the notations
\be
\delta_1 f(u,L,u^\dagger):= \frac{1}{2} \left(\delta_1 f^\ext(u,L,u^\dagger) +
\left(\delta_2 f^\ext(u,L,  u^\dagger)\right)^\dagger\right),
\label{derLi1}\ee
and
\be
Df(u,L, u^\dagger):= \frac{1}{2}\left(D f^\ext(u,L,u^\dagger) +
\left(D' f^\ext(u,L,  u^\dagger)\right)^\dagger\right).
\label{derLi2}\ee
These derivatives of $f$ can be checked to be independent of the choice of the extended function.

The Poisson bracket on $C^\infty(\cM_\herm)$ given in \cite{Li2} reads as follows:
\be
\{ f, h\}_{\cM_\herm} =- 2\langle \delta_1 f, D h\rangle_\R + 2\langle\delta_1 h, D f \rangle_\R - 2\langle R(u) Df, Dh\rangle_\R.
\label{LiPB2}\ee
This is evaluated at the arbitrary point $(u,L, u^\dagger)\in \cM_\herm$, and $R(u)\in \operatorname{End}(\cG^\C_\R)$ acts
non-trivially on the orthogonal complement of  $(\cT + \cA)$  according to
\be
R(u)X = - \left( \frac{1}{2}\coth\frac{1}{2} \ad_u\right)(X),
\quad \forall X\in (\cT^\perp + \cA^\perp),
\label{RLi}\ee
where we use $\cT^\perp$ and $\cA^\perp$ given in (\ref{I6}).

Now,  we introduce a one-to-one correspondence between the functions $f\in C^\infty(\cM_\herm)$ and
the functions $F\in C^\infty(\cH^o\times \fH)$ by the definition
\be
F(w,L):= f( 2w, L, 2 w^\dagger).
\label{Ff}\ee
The factor two is included for convenience (cf.~the definitions (\ref{RLi}) and (\ref{R})).
Since $\fH$ is an open submanifold of $\ri \cG$, we can take the derivatives
$\nabla_1F\in \cH$ and $\nabla_2F\in \cG^\C_\R$ similarly to (\ref{I10}).
The following simple statement is crucial for us.

\medskip \noindent
{\bf Lemma B1.}
\emph{The derivatives of $f\in C^\infty(\cM_\herm)$ given by (\ref{derLi1}) and (\ref{derLi2})  are related to the
derivatives of $F\in C^\infty(\cH^o\times \fH)$ by the identities
\be
Df(2w,L,2 w^\dagger) = L \nabla_2 F(w,L), \quad \delta_1 f(2w,L, 2w^\dagger) =\frac{1}{4} \nabla_1 F(w,L).
\label{B10}\ee}
\begin{proof}
For any $X\in \cG$, $Y\in \ri \cG$ and real parameter $t$,
the curves $e^{tX} L e^{-tX}$ and $ e^{tY} L e^{tY}$ stay in $\fH$, and
thus we have
\bea
&&F(w, e^{tX} L e^{-tX}) = f^\ext(2w, e^{tX} L e^{-tX}, 2 w^\dagger ), \nonumber\\
&&F(w, e^{tY} L e^{tY})= f^\ext(2w, e^{tY} L e^{tY}, 2 w^\dagger ).
\label{B11}\eea
Taking the derivative of the first identity gives
\be
\langle X L - L X, \nabla_2 F \rangle_\R  = \langle X, Df^\ext - D' f^\ext \rangle_\R,
\label{B12}\ee
while the second one gives
\be
\langle Y L + LY,  \nabla_2 F \rangle_\R = \langle Y, Df^\ext + D' f^\ext \rangle_\R,
\label{B13}\ee
where the arguments are as for $t=0$ in (\ref{B11}).
The relation (\ref{B12}) is readily seen to imply
\be
2 \langle X, L \nabla_2 F \rangle_\R = \langle X, Df^\ext + (D' f^\ext)^\dagger \rangle_\R,
\ee
and (\ref{B13}) implies
\be
2 \langle Y, L \nabla_2 F \rangle_\R = \langle Y, Df^\ext + (D' f^\ext)^\dagger \rangle_\R.
\ee
By combining these, we obtain the first equality in (\ref{B10}).

To obtain the second equality, we note that $f^\ext$ can be chosen to be independent of $v$, i.e.,
in such a way hat
\be
f^\ext( 2w, L, v) = f(2w, L) = F(w,L).
\ee
Then $(\delta_1 f) (2w,L)= \frac{1}{2} (\delta_1 f^\ext)(2w, L, 2w ^\dagger)$ follows from (\ref{derLi1}).
To continue, notice that
\be
f^\ext ( 2w + 2 t \xi, L,v) = F(w + t \xi, L), \quad \forall \xi\in \cH,
\ee
entails
\be
 \langle  2 \xi, (\delta_1 f^\ext) (2w, L, 2 w^\dagger)\rangle_\R =\langle \xi, \nabla_1 F(w,L) \rangle_\R.
\ee
 Consequently, we get
 \be
 \delta_1 f(2w,L) = \frac{1}{2} \delta_1 f^\ext(2w,L, 2 w^\dagger) = \frac{1}{4} \nabla_1 F(w,L),
 \ee
 as claimed. \end{proof}

The final result of the appendix is a direct consequence of the relations given by (\ref{B10}).

\medskip \noindent
{\bf Proposition B2.}
\emph{Via the correspondence (\ref{Ff}), the Poisson bracket (\ref{LiPB2}) satisfies
\be
\{ f,h\}_{\cM_\herm}(2w, L, 2 w^\dagger) = -\frac{1}{2} \{ F, H\}_\Li(w,L),
\ee
where $\{F,H\}_\Li$ is defined in (\ref{LiPB}).  Thus, up to an irrelevant overall constant,
which is due to conventions and the change of variable $u = 2w$,
$\{\ ,\ \}_{\cM_\herm}$ (\ref{LiPB2}) can be identified as the restriction of
$\{\ ,\ \}_\Li$ (\ref{LiPB}) to the dense open submanifold $\cH^o \times \fH \subset \cH^o \times \ri \cG$.}


\begin{thebibliography}{99}

\bibitem{Bar}
C. Bartocci, G. Falqui, I. Mencattini, G. Ortenzi and M. Pedroni, {\it On the geometric
origin of the bi-Hamiltonian structure of the
Calogero--Moser system}, Int. Math. Res. Not.  {\bf 2010} 279-296;
{\tt arXiv:0902.0953 [math-ph]}

 \bibitem{BH}
 H.W. Braden and N.W. Hone,
 {\it Affine Toda solitons and systems of Calogero--Moser type},
Phys. Lett. B {\bf 380} (1996) 296-302;
 {\tt arXiv:hep-th/9603178}

 \bibitem{CF2}
O. Chalykh and M. Fairon,
{\it On the Hamiltonian formulation of the trigonometric spin Ruijsenaars--Schneider system},
{\tt arXiv:1811.08727 [math-ph]}

 \bibitem{EV}
 P. Etingof and A. Varchenko,
 {\it  Geometry and classification of solutions of the classical dynamical Yang--Baxter equation},
Commun. Math. Phys. {\bf 192} (1998) 77-120;
 {\tt arXiv:q-alg/9703040}

\bibitem{Fa}
M. Fairon,
{\it Spin versions of the complex trigonometric Ruijsenaars--Schneider model from cyclic quivers},
Journ. Int. Syst. {\bf 4} (2019) xyz008;
{\tt arXiv:1811.08717 [math-ph]}

\bibitem{FMP}
G. Falqui, F. Magri and M. Pedroni,
{\it Bihamiltonian geometry, Darboux coverings, and linearization of the KP hierarchy},
Commun. Math. Phys. {\bf 197} (1998) 303-324;  	{\tt arXiv:solv-int/9806002}

\bibitem{FaMe}
G. Falqui and I. Mencattini,
{\it Bi-Hamiltonian geometry and canonical spectral coordinates for
the rational Calogero--Moser system}, J. Geom. Phys. {\bf 118} (2017) 126-137;
{\tt arXiv:1511.06339 [math-ph]}

\bibitem{F}
L. Feh\'er, {\it Poisson--Lie analogues of spin Sutherland models},
{\tt arXiv:1809.01529 [math-ph]}

\bibitem{FK}
L. Feh\'er and C. Klim\v c\'\i k,
{\it On the duality between the hyperbolic Sutherland and the rational
Ruijsenaars--Schneider models},
 J. Phys. A: Math. Theor. 42 (2009) 185202; {\tt arXiv:0901.1983 [math-ph]}

\bibitem{FP1}
L. Feh\'er and B.G. Pusztai,
{\it Spin Calogero models obtained from dynamical r-matrices and geodesic motion},
Nucl. Phys. B {\bf 734} (2006) 304-325;
{\tt arXiv:math-ph/0507062}

\bibitem{FP2}
L. Feh\'er and B.G. Pusztai,
 {\it Spin Calogero models associated with Riemannian symmetric spaces of negative curvature},
 Nucl. Phys. B {\bf 751} (2006) 436-458;  {\tt arXiv:math-ph/0604073}

\bibitem{GH}
J. Gibbons and T. Hermsen,
{\it A generalisation of the Calogero--Moser system},
Physica D {\bf 11} (1984) 337-348

\bibitem{GG}
T.F. G\"orbe and \'A. Gyenge,
{\it Canonical spectral coordinates for the Calogero--Moser space associated with the cyclic quiver},
{\tt arXiv:1812.02544 [math-ph]}

\bibitem{HT}
M. Henneaux and C. Teitelboim,
Quantization of Gauge Systems, Princeton University Press, 1992

 \bibitem{ILLS}
M. Isachenkov, P. Liendo, Y. Linke and V. Schomerus,
{\it Calogero--Sutherland approach to defect blocks},
 JHEP10(2018)204;
{\tt arXiv:1806.09703 [hep-th]}

\bibitem{KLOZ}
 S. Kharchev, A. Levin, M. Olshanetsky and  A. Zotov,
 {\it Quasi-compact Higgs bundles and Calogero--Sutherland systems with two types spins},
 J. Math. Phys. {\bf 59} (2018) 103509;
 {\tt arXiv:1712.08851 [math-ph]}

\bibitem{KZ}
 I. Krichever and A. Zabrodin,
{\it Spin generalization of the Ruijsenaars--Schneider model, non-abelian 2D
Toda chain and representations of Sklyanin algebra},
Russian Math. Surveys {\bf 50} (1995) 1101-1150;
{\tt arXiv:hep-th/9505039}

\bibitem{Li1}
L.-C. Li,
{\it Coboundary dynamical Poisson groupoids and integrable systems},
Int. Math. Res. Not. {\bf 2003} 2725-2746;  {\tt arXiv:math-ph/0506027}

\bibitem{Li2}
L.-C. Li,
{\it Poisson involutions, spin Calogero--Moser systems associated
with symmetric Lie subalgebras and the symmetric space spin Ruijsenaars--Schneider models,}
Commun. Math. Phys. {\bf 265} (2006) 333-372;
{\tt arXiv:math-ph/0506025}

\bibitem{LX}
L.-C. Li and P. Xu,
{\it A class of integrable spin Calogero--Moser systems},
Commun. Math. Phys. {\bf 231} (2002) 257-286;
{\tt arXiv:math/0105162 [math.QA]}

\bibitem{OR}
J.-P. Ortega and T. Ratiu,
Momentum Maps and Hamiltonian Reduction, Birh\"auser, 2004

 \bibitem{Res1}
N. Reshetikhin,
{\it Degenerate integrability of spin Calogero--Moser systems and the
duality with the spin Ruijsenaars systems},
Lett. Math. Phys. {\bf 63} (2003) 55-71;
 {\tt arXiv:math/0202245 [math.QA]}

\bibitem{Res}
 N. Reshetikhin,
{\it Degenerate integrability of quantum spin Calogero--Moser systems},
Lett. Math. Phys. {\bf 107} (2017) 187-200;
 {\tt arXiv:1510.00492 [math-ph]}

\bibitem{RS}
S.N.M. Ruijsenaars and H. Schneider,
{\it A new class of integrable systems and its relation to solitons},
Ann. Phys. {\bf 170} (1986) 370-405

 \bibitem{SS}
  V. Schomerus and E. Sobko,
 {\it From spinning conformal blocks to matrix Calogero--Sutherland models},
 JHEP04(2018)052;
 {\tt arXiv:1711.02022 [hep-th]}

\bibitem{Se}
A. Sergyeyev,
{\it A simple way of making a Hamiltonian system into a bi-Hamiltonian one},
Acta Appl. Math. {\bf 83} (2004)  183-197; {\tt arXiv:nlin/0310012 [nlin.SI]}

\bibitem{SL}
R. Sjamaar and E. Lerman,
{\it Stratified symplectic spaces and reduction},
Ann. of Math.  {\bf 134} (1991) 375-422

\bibitem{Sm}
R.G. Smirnov,
{\it Bi-Hamiltonian formalism: A constructive approach},
Lett. Math. Phys. {\bf 41} (1997)  333-347

\bibitem{Sn}
J. \'Sniatycki,
Differential Geometry of Singular Spaces and Reduction of Symmetries,
Cambridge University Press,  2013

\end{thebibliography}
\end{document}